\documentstyle{article}
\newcommand{\bra}[1]{\langle{#1}|}
\newcommand{\ket}[1]{|{#1}\rangle}
\newcommand{\braket}[2]{\langle{#1}|{#2}\rangle}
\renewcommand{\a}{\alpha}
\renewcommand{\b}{\beta}

\def\gh{{\rm gh}}
\def\sgh{{\rm sgh}}
\def\NS{{\rm NS}}
\def\R{{\rm R}}
\def\ii{{\rm i}}
\def\bz{{\bar z}}

\def\acomm#1#2{\left\{ #1, #2\right\}}
\def\tr{{\rm tr\,}}
\newlength{\bredde}
\def\slash#1{\settowidth{\bredde}{$#1$}\ifmmode\,\raisebox{.15ex}{/}
\hspace*{-\bredde} #1\else$\,\raisebox{.15ex}{/}\hspace*{-\bredde}  #1$\fi}
\newcommand{\NP}[1]{Nucl.\ Phys.\ {\bf #1}}
\newcommand{\PL}[1]{Phys.\ Lett.\ {\bf #1}}
\begin{document}
\rightline{NEIP-98-020}
\rightline{KUL-TF-98/59}
\rightline{\hfill December 1998}
\vskip 0.8cm
\centerline{\Large \bf Ramond-Ramond (boundary) states\footnote{%
Talk presented by M. Bill\'o at the 2nd 
Conference on Quantum Aspects of Gauge  Theories, 
Supersymmetry and Unification, Corfu, Greece, 21-26 September 1998.
Work partially supported by the European Commission
TMR programme ERBFMRX-CT96-0045.}}
\vskip 1.2cm
\centerline{\bf M. Bill\'o}
\vskip 0.1cm
\centerline{\sl Instituut voor theoretische fysica,}
\centerline{\sl K.U. Leuven, Celestijnenlaan 200D, B3001 Leuven, Belgium}
\centerline{\tt Marco.Billo@fys.kuleuven.ac.be}
\vskip 0.3cm
\centerline{\bf R. Russo}
\vskip 0.1cm
\centerline{\sl Institut de Phisique, Rue A.-L. Breguet 1 2000 
Neuch\^atel, Switzerland}
\centerline{\tt Rodolfo.Russo@iph.unine.ch}
\vskip 1cm
\begin{abstract}
The description of D-branes as boundary states for type II string  
theories (in the covariant formulation) requires particular care in the R-R  
sector. Also the vertices for R-R potentials that can couple to D-branes 
need a careful handling. As an illustration of this, the example of the 
D0-D8 system is reviewed, where a ``microscopic'' description of the 
interaction via exchange of R-R potentials becomes possible.
\end{abstract}
\section{Covariant superstring boundary states}
\label{secone}
D-branes can be incorporated in type II superstring theories by means  
of ``boundary states'' (b.s.) \cite{bs}. 
Such a state represents the insertion (at $\tau=0$) of a boundary
in the string world-sheet which lies within the
D-brane world-volume from the target-space point of view.
The b.s. formalism has been efficiently used to compute the  
interactions between moving D-branes \cite{BCDV} even in the presence  
of external fields \cite{06} \footnote{
See also the talks \cite{talks} of M. Bertolini and M. 
Serone at this conference, and references therein, for other  
applications of the b.s. formalism.}.
Here, following \cite{D0D8}, we will focus on some subtle aspects of the 
covariant b.s., which are relevant for the (anomalous) R-R couplings  
of D-branes  and for ``non-trivial'' systems, like for instance the D0-D8 one 
\cite{08sys,holiwu}.
\par
The boundary state $\ket B$ satisfies conditions that
correspond, in the closed string channel, to the boundary conditions  
for open strings ending on the D-brane. In the simple case of a flat  
D-brane located at $y^i=0$, the open string Neumann (Dirichlet) b.c.'s 
in the world-volume (transverse) coordinates $X^\alpha$ ($X^i$)  
translate into the conditions
$\partial_\tau X^\alpha |_{\tau=0}\ket{B_X} =  
X^i|_{\tau=0}\ket{B_X}=0$.
Expanding the closed string fields into oscillators, these conditions
become $(a^\mu_n + S^\mu_{~\nu}{\tilde a}^\nu_{-n})\ket {B_X}=0$ 
for $n\not = 0$ and
$\hat p^\alpha \ket{B_X} = \hat q^i \ket{B_X} = 0$ for the zero modes  
(we have introduced $S^{\mu\nu} =   (\eta^{\alpha\beta},-\delta^{ij})$).
The solution to these conditions reads
\begin{equation}
\label{capro1}
\ket{B_X} = \delta^{d_\perp}(\hat q)\, {\rm e}^{-\sum_{n=1}^\infty
a_{n}^\dagger\cdot S\cdot {\tilde a}^\dagger_{n}}\ket 0~.
\end{equation}
Analogous ``overlap'' conditions, {\it {i.e.}}
$(\psi -\ii \eta S\cdot{\tilde\psi})\ket{B_\psi,\eta}=0$ with $\eta=\pm
1$, arise for 
the fermionic fiels
$\psi^\mu,{\tilde\psi}^\mu$ fields whose mode expansion
depends on which sector (Ramond or Neveu-Schwarz) one considers.
Actually, the GSO projection requires a combination of the
two cases $\eta=\pm 1$, as we will see.
Imposing the BRST invariance of the b.s., $(Q + \tilde{Q})\ket B=0$, 
conditions for its ghost and superghost parts are obtained. 
Thus finally one determines explicitly the expression
\begin{equation}
\label{capro2}
\ket{B,\eta}_{\NS,\R} ={T_p\over 2} \ket{B_X}\ket{B_\gh}
\ket{B_\psi,\eta}_{\NS,\R} \ket{B_\sgh,\eta}_{\NS,\R}~.
\end{equation}
The D-brane tension, $T_p = \sqrt{\pi}(2\pi\alpha')^{3-p}$, is fixed 
factorising amplitudes involving boundary states \cite{turin}.
Each term in Eq. (\ref{capro2}) is expressed, like $\ket{B_X}$ in Eq.
(\ref{capro1}), by a coherent-state-type exponential of  
non-zero-modes and  a zero-mode part. In the NS sector, where there are no 
$\psi$ zero-modes, one has $\ket{B_\psi,\eta}_\NS$ 
$=\exp(\ii\eta\sum_{r=1/2}^\infty \psi_r^\dagger\cdot S\cdot  
{\tilde\psi}_r^\dagger)\ket0$.
In the following we will focus on the fermion and superghost 
zero-mode  part of the b.s. in the R-R sector. For complete  
expressions of the covariant b.s. see e.g. \cite{D0D8}. 
\subsection{The Ramond-Ramond sector}
\label{seconeone}
In the R sector, the $\psi$ field is integer moded and the
anti-commutations $\acomm{\psi_0^\mu}{\psi_0^\nu} = \eta^{\mu\nu}$  
imply that these zero-modes can be represented as $\Gamma^\mu$ matrices,
acting on a vacuum that carries spinor indices:
$\ket{A} \widetilde{\ket{B}}$. The zero-mode part of the b.s., 
satisfying the correct overlap conditions, turns out to be
\begin{equation}
\ket{B_\psi,\eta}_\R^{(0)} = {\cal M}^{(\eta)}_{AB} \ket A\ket  
{\tilde B}~,
\hskip 0.8cm\mbox{with}~~
{\cal M}^{(\eta)} = C\prod_\alpha \Gamma^\alpha \,\left(
\frac{1+\ii\eta\Gamma_{11}}{1+\ii\eta}\right)~,
\end{equation}
where $C$ is the charge conjugation matrix. For the superghost systems 
it is necessary to choose a picture $(P,\tilde P)$; we consider thus
a superghost vacuum $\ket{P,\tilde P}$ such that (e.g. in the l.m.  
sector)
$\gamma_m \ket P =  0$ for $m>P+1/ 2$ and 
$\beta_m \ket P =  0$ for $m> -P-3/2$.
\par
Insisting that the b.s. inserts a  boundary in the world-sheet, thus turning
e.g. a closed string sphere amplitude into an open string disk  
amplitude, implies that $P +\tilde P=-2$, due to the sphere and disk 
superghost anomaly\footnote%
{Consider a closed string sphere amplitude 
$\bra{0}\prod_i V_i(z,\bar z)\ket{B}$.
To cancel the sphere superghost number anomaly, the  superghost charges of
$\prod_i V_i(z,\bar z)$ must be $(-2-P,-2-\tilde P)$. If this  
amplitude has to coincide with the open string disk amplitude 
$\bra{0} \prod_i V_i(z) V_i(\bar z) \ket{0}$, the (open) 
superghost charge of the vertices must cancel the disk anomaly:
$-2-P -2 -\tilde P=-2$.}.
Due to the moding of the superghost fields, $P$ is integer in the
NS and half-integer in the R sector. Thus, the
boundary state $\ket{B}_{\R}$ has always $P\not=\tilde P$, and
can couple only to R-R states in a left-right asymmetric picture.
This point is crucial for our discussion. Let us choose for  
definiteness $P=-1/2, \tilde P = -3/2$; then
the superghost zero-mode part of the b.s. is 
\begin{equation}
\label{bsrsg0}
\ket{B_\sgh,\eta}_\R^{(0)} =
\exp\left[\ii\eta\gamma_0\tilde\beta_0\right]\,
  \ket{P=-{1/ 2},\tilde P=-{3/ 2}}~.
\end{equation}
We need to perform on the b.s. the relevant type IIA or B GSO projection, 
obtaining 
\begin{equation}
\label{bs22ba}
\ket{B}_\R  \equiv  {1 +(-1)^{p} (-1)^{F+G}\over 2}\,\,
{1 - (-1)^{\tilde F+\tilde G}\over 2}\, \ket{B,+}_\R =
{1\over 2} \Big( \ket{B,+}_\R + \ket{B,-}_\R\Big)~,
\end{equation}
where $p$ is even for Type IIA and odd for Type IIB. Notice that
the fermion and superghost number operators reduce in the
zero-mode sector as follows:
$(-1)^F\to\Gamma_{11}$ and ${G} \to - \gamma_0 \beta_0$.
\par
For later convenience we now rewrite the boundary state $\ket{B}_\R$
using 16-dimensional chiral and antichiral spinor indices $\a$ and
$\dot\a$ for Majorana-Weyl fermions. Then, for the Type IIA theory
we have
\begin{eqnarray}
\ket{B}_\R &=& \frac{T_p}{2}
\ket{B_X}\, \ket{B_\gh}\,\left\{
\left(C\Gamma^0\Gamma^{l_1}\ldots\Gamma^{l_p}\right)_{\a\b}
\cos\left[\gamma_0\tilde\beta_0+
\Theta\right]
\ket{\a}_{-1/2}\ket{\tilde \b}_{-3/2}\right.
\nonumber \\
&&+\left.
\left(C\Gamma^0\Gamma^{l_1}\ldots\Gamma^{l_p}\right)_{\dot \a\dot \b}
\sin\left[\gamma_0\tilde\beta_0 + \Theta\right]
\ket{\dot\a}_{-1/2}\ket{\tilde{ \dot\b}}_{-3/2}\right\}~~,
\label{cossin}
\end{eqnarray}
where $\Theta$ contains the non-zero modes \cite{D0D8},
and $\ket{\alpha}_P = \lim_{z\to 0} S^\alpha(z)\ket{P}$.
\section{Ramond-Ramond gauge fields}
\label{sectwo}
Let us focus on the R-R massless states of the type IIA theory.
The R-R states that are usually considered in the literature are
in the symmetric $(-1/2,-1/2)$ picture and are created by
the following vertex operators:
\begin{equation}
V_\R(k;z,\bz) = \frac{1}{2\sqrt{2}}\,
(C{ F}^{\,(m+1)})_{\alpha{\dot{\beta}}}
\, {V}^\a_{-1/2}(k/2;z) \,
\tilde{V}^{\dot \b}_{-1/2}(k/2;\bz)~~,
\label{simver}
\end{equation}
where 
\begin{equation}
(C{ F}^{\,(m+1)})_{\alpha{\dot{\beta}}} =
\frac{\left( C \Gamma^{{\mu_1} \dots
{\mu}_{m+1}} \right)_{\alpha{\dot{\beta}}}}{ (m+1)!} \,
F_{\mu_1 \dots \mu_{m+1}}
\label{simver1}
\end{equation}
with $m$ odd, and
${V}^\a_{\ell}(k;z) = c(z)\,S^{\a}(z)\,{\rm
e}^{\ell\,\phi(z)} \,{\rm e}^{\ii k\cdot X(z)}$.
The form ${F}^{(m+1)}$ has the right degree to be interpreted as a 
R-R field strength of the type IIA theory; indeed
the vertex $V_\R$ is BRST invariant only if $k^2=0$, and
$d{ F}^{(m+1)} =d\,*{ F}^{(m+1)}=0$ which are precisely the Bianchi  
and Maxwell equations of a field strength.
Thus, from the target space point of view, it is clear that D-branes  
can not emit the states of Eq.~(\ref{simver}); in fact, they are 
charged object that should be minimally coupled to the R-R fields and  
so their vertices must contain directly the gauge potential.
However, also from the world-sheet point of view, it is clear that  
the boundary state can not be saturated with the usual R-R states.
In fact, $\ket{B}_{\R}$ is in the $(-1/2,-3/2)$ picture of the R-R
sector, and thus, to soak up the superghost number anomaly, it can  
only couple to states that are in the asymmetric $(-3/2,-1/2)$ picture. 
These two observations are intimately related; in fact by considering 
the asymmetric picture \cite{D0D8}, it is possible to construct 
vertices that are equivalent to those of Eq.~(\ref{simver}), but that 
contain gauge potentials instead of field strengths.
They are expressed as the sum of infinite terms:
\begin{equation}
W(k;z,\bz) = \sum_{M=0}^\infty W^{(M)}(k;z,\bz)~.
\label{Rvertex}
\end{equation}
Each term contains the same polarisation tensor $A^{(m)}_{\mu_1\ldots\mu_m}$ 
with the right degree for a $m$-form potential. Explicitly the first two are
\begin{eqnarray}
\label{iproc2}
W^{(0)}(k;z,\bz) & = &
(C{ A}^{\,(m)})_{\alpha{{\beta}}}
\, {V}^\a_{-1/2}(k/2;z) \,
\tilde{V}^{\b}_{-3/2}(k/2;\bz)~,\nonumber\\
W^{(1)}(k;z,\bz) & = &- (C{ A}^{\,(m)})_{\dot\a \dot\b}
~\eta(z) {V}^{\dot\a}_{+1/2}(k/2;z) ~
\bar\partial{\tilde \xi}(\bz)
\tilde{V}^{\dot\b}_{-5/2}(k/2;\bz)~,
\end{eqnarray}
where $CA^{(m)}$ is s given by an expression similar to Eq. (\ref{simver1}).
All subsequent term can be determined by requiring BRST invariance  
under the total charge $Q + \tilde{Q}$. 
\par
The invariance is obtained non-trivially. Recall that there is a  
natural splitting of the BRST charge ${Q}$ into ${Q}_0$ (corresponding  
roughly to a current $j_0(z)\propto c(z) T(z)$, $T(z)$ being the  
stress-energy tensor), ${Q}_1$ (with current roughly 
$j_1(z)\propto \gamma(z) G(z)$,  $G(z)$ being the supercurrent) and 
${Q}_2$ (containing only ghosts and antighosts, needed for the closure).
The commutation with ${Q}_0$ (and $\tilde{Q}_0$) requires the
conformal weight of the vertex to be zero, i.e. simply that
$k^2=0$. The commutation with ${Q}_2$ can be arranged 
for all terms without imposing any condition. 
\par
The important part is the commutation with ${Q}_1 + \tilde{Q}_1$.
We have $[\tilde{Q}_1,W^{(0)}]=0$ and it turns out that for all
the subsequent terms
$[{Q}_1,{W^{(M)}}] +$ $[\tilde{Q}_1,{W^{(M+1)}}]=0$
provided the transversality condition for the polarisation is  
satisfied:
\begin{equation}
\slash{k}{ A}^{(m)}+{ A}^{(m)}\slash{k} = 0 ~~\Rightarrow d*A^{(m)}~.
\label{lorentz}
\end{equation}
This is the usual gauge condition for a potential. 
Notice that the constraint in Eq.~(\ref{lorentz}) is {\em  independent} 
of the on-shellness condition $k^2=0$. 
Thus the asymmetric R-R vertices share the property of the usual  NS-NS 
massless vertices that they can be extended ``softly'' off-shell,  
namely spoiling only the commutation with the $Q_0$ ($\tilde Q_0$) 
part of the BRST charge\footnote{%
This off-shell extension is related to the scheme proposed in  
\cite{RT} as the off-shell vertices still belong to the ``restricted'' 
cohomology $Q'+\tilde Q'$ introduced there.}.
It is argued in \cite{D0D8} that from such an off-shell extension  
sensible result are obtained in the field theory limit when considering,
as in \cite{class}, the emission of closed strings from D-branes, 
where one is forced to go off-shell.
\par
The state created by  the vertex operator (\ref{Rvertex}) 
has a rather simple expression when written in the $(\beta,\gamma)$  
system:
\begin{eqnarray}
\label{-3/2state}
\ket{W} &=&
\left(C{ A}^{(m)}\right)_{\a\b}
\cos\left(\gamma_0{\tilde \b}_0\right)
\ket{\a;k/2}_{-1/2}\,\ket{{\tilde \b};k/2}_{-3/2}
\nonumber\\
&&+\left(C{ A}^{(m)}\right)_{\dot\a \dot\b}
\sin\left(\gamma_0{\tilde \b}_0\right)
\ket{\dot\a;k/2}_{-1/2}\,\ket{{\tilde {\dot\b}};k/2}_{-3/2}~,
\end{eqnarray}
where we have introduced the notation
$\ket{\alpha;k}_{\ell}
\equiv \lim_{z\to 0} {V}^\a_\ell(k;z)~\ket{0}$.
The state $\ket{W}$ is similar in form to the zero-mode part
of the boundary state $\ket{B}_\R$, see Eq. (\ref{cossin}).
\subsection{Regulated scalar products and amplitudes} 
\label{sectwoone}
Having constructed the D-brane boundary states, and the 
closed string states that can couple to them, it is possible for  
instance to compute interactions between branes, 
$\bra{B_1} D \ket{B_2}$  ($D$ being the closed string propagator),
closed string fields emission from the D-brane, $\bra{B}D\ket{W}$,  
where $\ket{W}$ represents the appropriate (off-shell) state, as well as 
to compute the scalar products in the Hilbert space of these states: 
$\braket{W^1}{W^2}$.  
\par
In all these cases an important subtlety is encountered in the R-R
sector. Indeed the superghost zero-modes give rise to an infinite or 
ill-defined expression, which we need to properly regulate.
The prescription we introduce (and that was devised for the purely  
Neumann boundary states in \cite{YOST}) consists in defining the  
``dangerous''
expressions as  the limit for $x\to 1$ of the expressions modified by the
introduction of the ``regulator'' 
\begin{equation}
\label{reg1}
{\cal R}(x)\equiv x^{2(F_0 + G_0)}~.
\end{equation}
$F_0$ and $G_0$ are the zero-mode part of the fermion and superghost  
numbers respectively.
\par
Since $G_0 = -\gamma_0 \beta_0$, one can immediately
compute the regulated superghost part. On the other side,
one usually only considers $(-1)^{F_0} = \Gamma^{11}$. To give a meaning to
$F_0$ and thus to $x^{2 F_0}$, let us group the directions $\mu$ into  
pairs. A two-state Hilbert space is associated to each pair.
For instance, in the directions $(1,2)$ the fermionic creator-annihilator 
couple is $e^\pm_1 = (\Gamma^1 \pm\ii \Gamma^2)/2$, and the space is  
spanned by $\ket{\uparrow}$ and $\ket{\downarrow}$, with eigenvalues $+1$ and  
$-1$ resp. of the number operator $N_1 = -\ii\Gamma^{12}$.
Notwithstanding the explicit breaking of SO$(10)$ to SO$(2)^5$,
if all directions were  equivalent the final result would be still  
SO$(10)$ invariant, see \cite{YOST}. However, in presence of D-branes, care is  
needed in the pairing to obtain meaningful results, see \cite{D0D8}; see  
also Sec. (\ref{secthreeone}).
In this framework we find (after Euclidean rotation) that
\begin{equation}
\label{capro3}
(-1)^{F_0} = \Gamma^{11} = \ii\prod_{k=1}^5 N_k = \prod_{k=1}^5 
\exp(\ii N_k{\pi\over 2}) = (-)^{{1\over 2} \sum_{k=1}^5 N_k}~.
\end{equation}
Thus $F_0 = {1\over 2} \sum_{k=1}^5 N_k$ and finally
$x^{2F_0}= x^{\sum_{k=1}^5 N_k}$. We can now explicitly compute 
regulated expressions\footnote{%
Typically we have to compute expressions like 
$\tr (x^{2F_0}\prod_I \Gamma^I)$ or 
$\tr (x^{2F_0}\prod_I \Gamma^I \Gamma^{11})$,
that give rise in the various subspaces, depending on the set of indices 
$\{I\}$ and the chosen decomposition in pairs, to factors of 
$\tr(x^{N_k})= (x+1/x)$ or of 
$\tr (\ii x^{N_k} N_k) = \ii (x -1/x)$.}.  
Notice that, because of Eq.~\ref{reg1}, the 
fields $\psi$ and the superghosts are not any more factorised; 
the regulator cancels the ill-defined 
contribution of the superghost fields with the one coming from two 
$\psi$'s, leaving a well-defined result that matches 
the one found in the light cone computations.
\subsection{Null states and ``duality''  relations among R-R  
potentials}
\label{sectwotwo}
A remarkable difference between the usual states and those of
Eq.~(\ref{-3/2state}) is that the first ones have a definite  
chirality,
while the second ones mix dotted and undotted indices. Thus one may  
wonder whether also for the asymmetric states it is possible to derive the
Hodge conditions on the potentials that correspond to the usual
constraints $F^{(p+2)}$ = $*F^{(8-p)}$. As we will see, this
feature arise in a non-trivial way from the Hilbert space structure  
of the new states.
\par
The scalar product between a bra and a ket state 
for the R-R massless fields  has to be defined as
$\langle W'\,,\,W\rangle \equiv
\lim_{x\to 1}\;\bra{W'}\,{\cal R}(x)\,\ket{W}$.
It is  easy to see that, with this prescription, the states
of Eq. (\ref{-3/2state}) have a definite norm.
Moreover with this definition, the one-to-one mapping between asymmetric 
states and the usual ones becomes an isometry \cite{D0D8}.
The most striking feature of the regulated scalar product is that  
forms of different order are, in general, not orthogonal to each other. 
Then there is a degenerate metric and therefore null-states.
\par
For all states with transverse polarisations, the decoupling of the  
null states implements the dualities between R-R potentials. 
Consider a 1-form and a 7-form state with polarisations $A_1$ 
and $A_{2\ldots 8}$, transverse
and on-shell: $k_0=k_9$, other momenta zero. The Hodge duality
$F_{01} = F_{2\ldots 9}$ implies $A_1=A_{2\ldots 8}$. This equality
is enforced in our context by setting to zero the null vector
$\ket{W_1} - \ket{W_{2\ldots 8}}$ that arises because of the  
non-orthogonality of the two corresponding asymmetric states $\ket{W_1}$ and  
$\ket{W_{2\ldots 8}}$, given by Eq. (\ref{-3/2state}) with $CA^{(m)}$ 
equal to $C\Gamma^1$, resp. $C\Gamma^{1\ldots 8}$.
\par
For longitudinal and scalar states (unphysical, but the one involved  
in Coulomb-like interactions!) a similar phenomenon occurs.
Consider, for example, a 1-form state with polarisation $A_0$, off-shell 
in the kinematic region $k^i=0$, $i=0,1\ldots 8$,
relevant for the study of D-brane interactions as in Sec.
(\ref{secthreetwo}), and a 9-form state $A_{0\ldots 8}$ in the same  
region.
It is found that the corresponding states $\ket{W_0}$ and 
$\ket{W_{0\ldots 8}}$ have a degenerate metric admitting the null  
state 
\begin{equation}
\ket{\chi}=\ket{W_0}+\ket{W_{0\ldots8}}~~.
\label{null2}
\end{equation}
While on-shell the scalar and longitudinal polarisations like
always decouple from physical
amplitudes, the presence of a D-brane forces an off-shell
continuation and only linear combinations decouple. 
When we set $\ket{\chi}=0$, we are led to identify
$A_0$ with $-A_{0\ldots8}$, recovering in this way  an unusual  
``duality''
also off-shell.
\par
The Hilbert space structure is thus responsible both for the
identification $A_1=A_{2\ldots 8}$ for physical states, 
that is usually seen as an effect of
the GSO projection of the symmetrical states, and for the
`unusual'' identification $A_0 = -A_{0\ldots 8}$, 
that is proper only of the longitudinal asymmetric states
(\ref{-3/2state}).
\section{D0-D8 interaction: ``microscopic'' interpretation} 
\label{secthree}
We consider now the system of a D0 and a D8 brane at rest. This system
has been widely investigated from different point of views
\cite{08sys,holiwu}. One of its important features is the  
creation of a fundamental string when the two branes cross each 
other mantaining  the BPS condition. 
Although the description of this phenomenon, e.g. from the
(M)atrix theory point of view \cite{holiwu}, can be matched with the  
b.s. treatment, here we do not dwell on this aspect. We simply show how  
the peculiar properties of the asymmetric R-R states (\ref{-3/2state})
shed light on the interpretation of the R-R part of the D0-D8  
interaction.  
\subsection{D0-D8 interaction in the  boundary states formalism}
\label{secthreeone}
We take a D0-brane, with world-volume direction (0), and a D8-brane
(012345678), at distance $b$ in the transverse direction.
Since the number $\nu$ of ``mixed'' (ND + DN) directions is 8, such a 
configurations is BPS \cite{POLCH}, and we expect the amplitude to  
vanish.
\par
In the boundary state formalism, the amplitude ${\cal A} =  
\bra{B^8}D\ket{B^0}$
is straightforwardly computed in the NS-NS sector, with  the result
\begin{equation}
\label{ansns}
{\cal A}_{\rm NS}
= \frac{T}{2\pi}{1\over  \sqrt{8\pi^2\alpha'}}
\int_0^\infty{dt} 
\left(\pi\over t \right)^{1\over 2}\,{\rm e}^{-b^2 /(2\alpha' t)}\,
\left[ \left({f_4\over f_2}\right)^8 - \left({f_3\over f_2}\right)^8
\right]  = -{T\over 4\pi\alpha'}|b|~.
\end{equation}
$T$ is the ``volume'' of the common word-volume dimension, the time,
and the functions $f_i(q)$, with $q={\rm e}^{-t}$, are the usual
infinite products given in \cite{POLCH}.
In the first equality, the two terms within square brackets represent  
the oscillator contributions to the two part of the GSO projected  
amplitude, with or without the insertion of $(-)^{F+G}$. In the second step,
one uses the ``aequatio identica'' $f_4^8 - f_3^8 = f_2^8$ to replace  
1 for their sum and carries out explicitly the integral. The final result
represents a net repulsive Coulomb potential in the transverse  
direction, resulting from the exchange of graviton and dilaton, while the  
exchanges of massive fields cancel.
Considering the amplitude in the dual channel, i.e. as the 1-loop
free energy of the open strings suspended between the  
D-branes\footnote{%
Also in the open string set-up there is a GSO projection
and in the R sector one needs a
``regularisation'' to compare the behaviour of fermionic and
superghost zero-modes. In the R trace the superghosts
provide then (correctly) just  a factor 1/2.
In the R trace with $(-)^F$ inserted, which is essentially a Witten  
index, the contribution of all non-zero modes vanishes; however in the  
zero-mode sector, the regularisation provides a compensation  between a 0 from  
the fermionic zero-modes and an infinity from the superghost precisely  
when $\nu=8.$}, the two terms above correspond 
to the traces of ${\rm e}^{-2t' (L_0-a)}$ in the NS or 
R sector of the theory.
\par
The R-R contribution to the amplitude is more delicate. We need to  
regulate the
expression; to this effect we may split the directions into the pairs
(1,2), (3,4),\ldots, (9,0). It is essential (and rather intuitive)  
that 
one must not pair NN or DD directions with ``mixed'' ones, see
\cite{D0D8}.
Considering the amplitudes before GSO projection,
using the definitions of Sec. (\ref{secone}) one finds for the  
superghost zero-modes
\begin{equation}
\label{bs50}
{}_\R^{(0)}\!\langle B^8_\sgh,\eta | \, x^{2G_0} \,
|B^0_\sgh,\eta'\rangle_\R^{(0)} =
(1 - \eta_1 \eta_2 x^2)^{-1}~.
\end{equation}
For the fermionic zero-modes one one has to compute
\begin{equation}
\label{oproc1}
{}_\R^{(0)}\!\langle B^8_\psi\eta | \, x^{2F_0} \,
|B^0_\psi,\eta'\rangle_\R^{(0)} =
-\tr(x^{2F_0} \Gamma^0) \delta_{\eta \eta',-1} 
-\tr(x^{2F_0} \Gamma^0\Gamma^{11}) \delta_{\eta \eta',1}~.
\end{equation}
Employing the techniques devised in Sec. (\ref{sectwoone}) to treat  
Eq. (\ref{oproc1}) one finally obtains
\begin{equation}
\label{oproc2}
\lim_{x\to 1} 
{}_\R^{(0)}\!\langle B^1,\eta_1 | {\cal R}(x) |
B^2,\eta_2\rangle_\R^{(0)} = 16 \delta_{\eta \eta',1}~.
\end{equation}
Comparing with the GSO projected expression (\ref{bs22ba})
one sees that it is only the term {\em without} insertion of
$(-)^{F+G}$ (the ``odd spin structure'')
that contributes, contrarily to what happens for all cases with $\nu\not= 8$.
\par
It is easy to see that for the odd spin structure the contribution of  
all non-zero modes vanishes; taking into account the contribution of the  
bosonic zero-mode,which does not depend from the sector, we finally get
\begin{equation}
\label{arr}
{\cal A}_\R
= \frac{T}{2\pi}{1\over  \sqrt{8\pi^2\alpha'}}
\int_0^\infty{dt} 
\left(\pi\over t \right)^{1\over 2}\,{\rm e}^{-b^2 /(2\alpha' t)}
= {T\over 4\pi\alpha'}|b| = -{\cal A}_\NS~.
\end{equation}
\par
The computation of the R-R amplitude is not parity-invariant, due to
the $\Gamma^{11}$ in Eq. (\ref{oproc1}).
If we switch position between the two
branes, $x^9\to -x^9$, we get 
${\cal A}_{NS} - {\cal A}_R = -T |b|/(2\pi\alpha')$. 
The BPS zero-force condition is mantained if a fundamental string with 
tension $1/(2\pi\alpha')$ is created.
\par
\subsection{Interpretation in terms of field theory exchange  
diagrams}
\label{secthreetwo}
Thus, the repulsive NS contribution to the D0-D8 amplitude
is cancelled by an attractive Coulomb force in the ``odd spin structure''
(R-R sector); this is different from what happens
in the D0-D6 brane system where the odd-spin structure encodes the Lorentz
force \cite{06}.  
What kind of exchange can be responsible for this Coulomb 
interaction? The D8 brane naturally couples to the
R-R 9-form potential, and the D0 brane to the 1-form potential, and  
the two seem not to communicate.
However the forms emitted by the static D-branes have to be off-shell 
(they carry only transverse momentum). Moreover, whatever is
exchanged between the two can only have momentum along the 9th  
direction (as appropriate for the interpretation of the Coulomb
interaction as due to an exchange diagram).
Therefore the R-R states that may appear in the factorisation of the 
D0-D8 amplitude are precisely the ``unphysical'' states 
$\ket{W_{01\ldots 8}}$ and $\ket{W_0}$
discussed in Sec. (\ref{sectwotwo}) above. The peculiar ``duality''  
relation between them, i.e. the existence of the null vector $\ket\chi$ of 
Eq. (\ref{null2}) allows an interpretation of the R-R part of the 
D0-D8 interaction as the exchange of a state that can be described 
equivalently by the 1-form $A_0$ or by the 9-form $A_{0\ldots8}$. 
Namely, the Fourier transform of the R-R amplitude (\ref{arr})  
factorises as  follows: 
\begin{equation}
\label{fac1}
\tilde{\cal A}_\R(k_9) \propto
{}_\R\!\bra{B^8}(\ket{W_{01\ldots 8}} - \ket{W_0})\
{1\over k_9^2}
(\bra{W_{01\ldots 8}} - \bra{W_0})\ket{B^0}_\R~,
\end{equation}
which explains also the attractive character of this interaction, as opposed 
to the usual R-R repulsive force in the other static D-brane interactions.
\vskip 0.1cm\noindent
{\bf Acknowledgements} We thank P. Di Vecchia, M. Frau, A. Lerda,
and S. Sciuto for discussions and reading of the manuscript and 
M. Serone for discussions. 

\begin{thebibliography}{99}
%
\bibitem{POLCH} J. Polchinski, Phys. Rev. Lett {\bf 75} (1995) 4724,
{\tt hep-th/9510017}; J. Polchinski, S. Chaudhuri and C.V. Johnson,  
{\it
Notes on D-branes}, {\tt hep-th/9602052};
J. Polchinski, {\it TASI lectures on D-branes}, 
{\tt hep-th/9611050}.
%
\bibitem{bs}
C.G. Callan, C. Lovelace, C.R. Nappi and S.A. Yost,
\NP{B308} (1988) 221; J. Polchinski and T. Cai, \NP{B286} (1988) 91.
M.B. Green and M. Gutperle, \NP{B476} (1996) 484, {\tt  
hep-th/9604091}.
%
\bibitem{BCDV}M. Billo', P. Di Vecchia and D. Cangemi,
\PL{400B} (1997) 63, {\tt hep-th/9701190}.
%
\bibitem{06} M. Bertolini, R. Iengo and C. Scrucca, \NP{B522} (1998) 193,
{\tt hep-th/980110}.
M. Billo', P. Di Vecchia, M. Frau, A. Lerda, R. Russo 
and S. Sciuto, Mod. Phys. Lett. {\bf A13} (1998) 2977 {\tt hep-th/9805091}.
%
\bibitem{talks} M. Bertolini, P. Fr\'e, R. Iengo, and 
C. A. Scrucca, 
{\tt hep-th/9810150}; 
M. Serone, J.F. Morales, J.C. Plefka, C.A. Scrucca and  A.K.
Waldron, 
{\tt hep-th/9812039}. 
%
\bibitem{D0D8} M. Bill{\'{o}}, P. Di Vecchia, M. Frau, A. Lerda, I.  
Pesando, R. Russo and S. Sciuto, 
Nucl. Phys {\bf B526}(1998) 199, {\tt hep-th/9802088}.
%
\bibitem{08sys} C. P. Bachas, M. R. Douglas and  M. B. Green, 
JHEP {\bf 07} (1997) 002, {\tt hep-th/ 9705074};
O. Bergman, M. Gaberdiel and G. Lifschytz,
Nucl. Phys. {\bf B509} (1998) 194, {\tt hep-th/9705130};
%
U. Danielsson, G. Ferretti and I. R. Klebanov,
Phys. Rev. Lett. {\bf 79} (1997) 1984, \newline {\tt hep-th/9705084}.
%
\bibitem{holiwu} P. M. Ho, M. Li and Y. S. Wu, 
Nucl. Phys. {\bf B525} (1998) 146, {\tt hep-th/9706073};
P. M. Ho and Y. S. Wu, 
Phys. Lett. {\bf B420} (1998) 43, {\tt hep-th/9708137}.
%
\bibitem{turin} M. Frau, I. Pesando, S. Sciuto, A. Lerda and R.  
Russo,
\PL{400B} (1997) 52,\newline {\tt hep-th/9702037}.
%
%
\bibitem{RT} S. Ramgoolam and L. Thorlacius,
\NP{B483} (1997) 248, {\tt hep-th/ 9607113}.
%
\bibitem{class}P. Di Vecchia, M. Frau, I.
Pesando, S. Sciuto, A. Lerda, R. Russo, Nucl. Phys. {\bf B507} (1997)  
259, 
{\tt hep-th/9707068}.
%
\bibitem{YOST} S.A. Yost, Nucl. Phys. {\bf B321} (1989) 629.
%
\end{thebibliography}
\end{document}